\documentclass[]{spie}  

\usepackage{amsmath,amsfonts,amssymb}
\usepackage{graphicx}
\usepackage[colorlinks=true, allcolors=blue]{hyperref}

\title{Novel infrared-blocking aerogel scattering filters and their applications in astrophysical and planetary science observations}

\author[a,b]{Kyle R Helson}
\author[d]{Stefan Arseneau}
\author[b]{Alyssa Barlis}
\author[d]{Charles L. Bennett}
\author[b]{Thomas M. Essinger-Hileman}
\author[e]{Haiquan Guo}
\author[d]{Tobias Marriage}
\author[b]{Manuel A. Quijada}
\author[d]{Ariel E. Tokarz}
\author[c]{Stephanie L. Vivod}
\author[b]{Edward J. Wollack}

\affil[a]{The University of Maryland, Baltimore County. Baltimore, Maryland, USA}
\affil[b]{NASA Goddard Space Flight Center, Greenbelt, Maryland, USA}
\affil[c]{NASA Glenn Research Center, Cleveland, Ohio, USA}
\affil[d]{The Johns Hopkins University, Baltimore, Maryland, USA}
\affil[e]{Universities Space Research Association, Columbia, Maryland, USA}

\authorinfo{Further author information: (Send correspondence to Kyle Helson.)\\ Kyle Helson: E-mail: kyle.helson@nasa.gov}

\begin{document} 
\maketitle

\begin{abstract}
Infrared-blocking scattering aerogel filters have a broad range of potential applications in astrophysics and planetary science observations in the far-infrared, sub-millimeter, and microwave regimes. Successful dielectric modeling of aerogel filters allowed the fabrication of samples to meet the mechanical and science instrument requirements for several experiments, including the Sub-millimeter Solar Observation Lunar Volatiles Experiment (SSOLVE), the Cosmology Large Angular Scale Surveyor (CLASS), and the Experiment for Cryogenic Large-Aperture Intensity Mapping (EXCLAIM). Thermal multi-physics simulations of the filters predict their performance when integrated into a cryogenic receiver. Prototype filters have survived cryogenic cycling to 4\,K with no degradation in mechanical properties. 
\end{abstract}


\section{INTRODUCTION}
\label{sec:intro}  
Cryogenic telescope receivers common in the far-infrared, sub-millimeter, and microwave regimes require strong out-of-band infrared rejection in order to maintain cryogenic performance. A variety of different strategies have been employed to realize thermal blocking structures, which include reflective meshes, absorptive, and scattering designs\cite{ULRICH196737, ULRICH196765, ade_filters, Bock:95, Halpern, Munson:17, Timusk:81, choi, Whitbourn:85}. These low-pass filters have been implemented using a variety of different materials. 

We report on the continued development of broadband, tunable, polymer aerogel-based infrared-blocking filters for millimeter, sub-millimeter, and far infrared observations in planetary science and astrophysics, first reported in Ref. \citenum{Essinger-Hileman:20}. The filters presented are primarily polyimide aerogels\cite{guo, guo2, meador1} loaded with diamond scattering particles ranging from 1 to 60 microns, although different polymers for the host aerogel matrix have been explored. These filters have strong out-of-band rejection but also a low index of refraction, (n $\simeq$ 1.15) eliminating the need for anti-reflection (AR) coating. These filters function primarily by scattering away unwanted light and subsequently preventing it from reaching the coldest parts of the receiver. 

\section{Optical Modeling and Filter Design}
Filter optical design relies primarily on a Mie scattering dielectric model\cite{Essinger-Hileman:20}. Mie scattering occurs because of the contrast between the higher index of refraction of the embedded particles and the lower index of the surrounding medium. Table \ref{tab:materials_prop} shows a summary of material and optical properties of substrates used to construct aerogel scattering filters. While this publication focuses only on diamond-loaded polyimide aerogels, data for silica and silicon are provided for context. Even though bulk polyimide has an index of refraction of 1.7, due to the low volume filling fraction of an aerogel, an unloaded polyimide aerogel typically has an index of refraction of about 1.1. The diamond-loaded aerogel filters have an index of refraction typically less than 1.2. 

\begin{table}[h]
\centering
\begin{tabular}{|c|c|c|} 
 \hline
 \textbf{Material} & \textbf{Index of Refraction, n} &\textbf{Density [$\frac{g}{cc}$]} \\ 
 \hline
 Silica & 1.95 & 2.65 \\
 \hline\
 Polyimide & 1.7 & 1.42 \\
 \hline
 Silicon & 3.4 & 2.33 \\
 \hline
 Diamond & 2.38 & 3.53  \\
 \hline
\end{tabular}
\caption{Materials properties for bulk components of scattering aerogel filters. Early prototypes used a variety of materials, but the filters in this publication used only polyimide and diamond.}
\label{tab:materials_prop}
\end{table}

\begin{figure}[h]
    \centering
    \includegraphics[width = 0.52\textwidth]{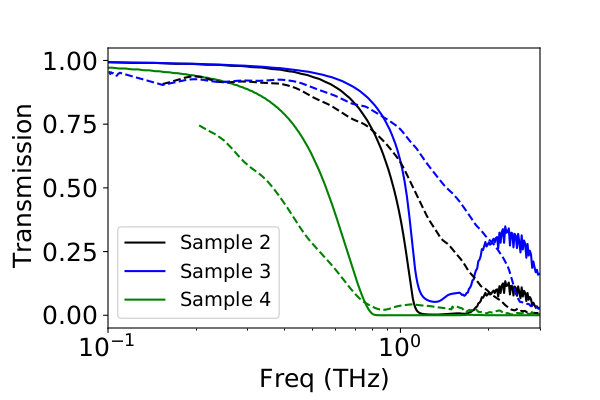}
    \includegraphics[width = 0.40\textwidth]{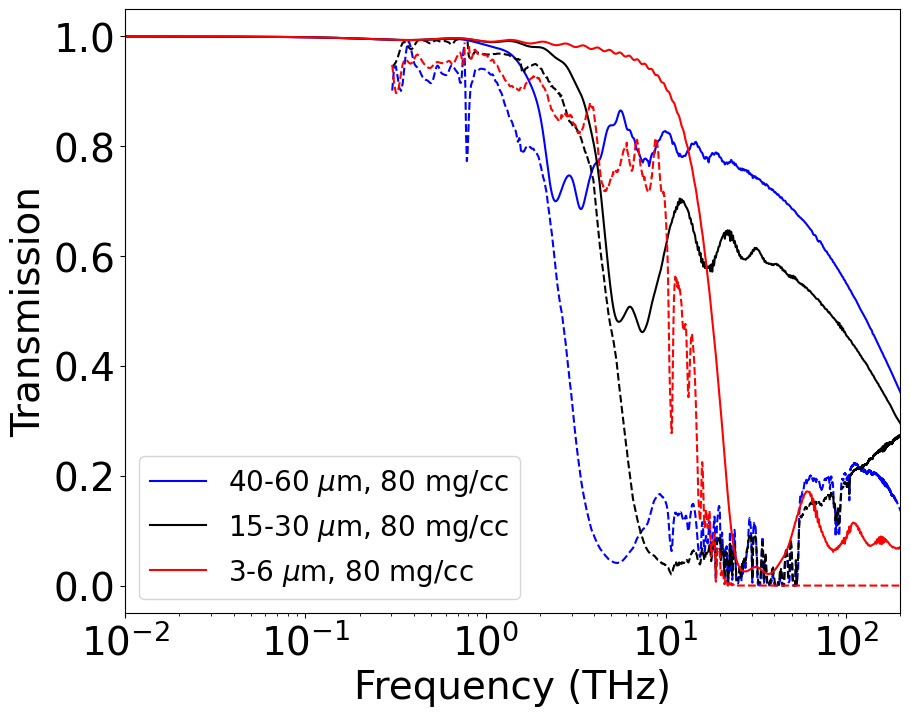}
    \caption{Comparison between early model versus measurement and current modeling versus measurement. \textit{Left:} Comparison of Mie scattering model and measured silica aerogels with silicon particles. Solid lines represent the aerogel dielectric model, while the dashed lines are measured transmission spectra. \textit{Right:} Comparison three different diamond-loaded polyimide aerogel sample transmission data (dashed lines) and their respective Mie scattering models (solid lines). Transmission data and model are scaled to represent 200\,$\mu$m thick films loaded at 80\,mg/cc. Blue curves are for a filter with 40-60\,$\mu$m particles, black curves are for a filter with 15-30\,$\mu$m particles, and the red curves are for a filter with 3-6\,$mu$m particles. While the very high frequency disagreement still persists, especially for larger particles, the current Mie scattering model predicts filter cutoff frequencies better than the early model.}
    \label{fig:mie_model}
\end{figure}

Early filter prototypes from Ref.~\citenum{Essinger-Hileman:20} were fabricated with silicon powder produced by grinding up float-zone silicon wafers in a ball mill. The resulting powder was sieved through meshes of various sizes, down to US 140 mesh size, or 106 microns. This produced a powder that had a broad, non-uniform particle size distribution. Early modeling efforts assumed this to be a uniform size distribution in particle radius, however later analysis showed a significant excess of smaller particles. The large dispersion in the particle size distribution produced by ball milling and subsequent sieving process is responsible for the relatively soft cut-off frequency in the measurement and the disagreement between measurement and model seen in the left plot of Figure \ref{fig:mie_model}. 

\begin{figure}[h]
    \centering
    \includegraphics[width = \textwidth]{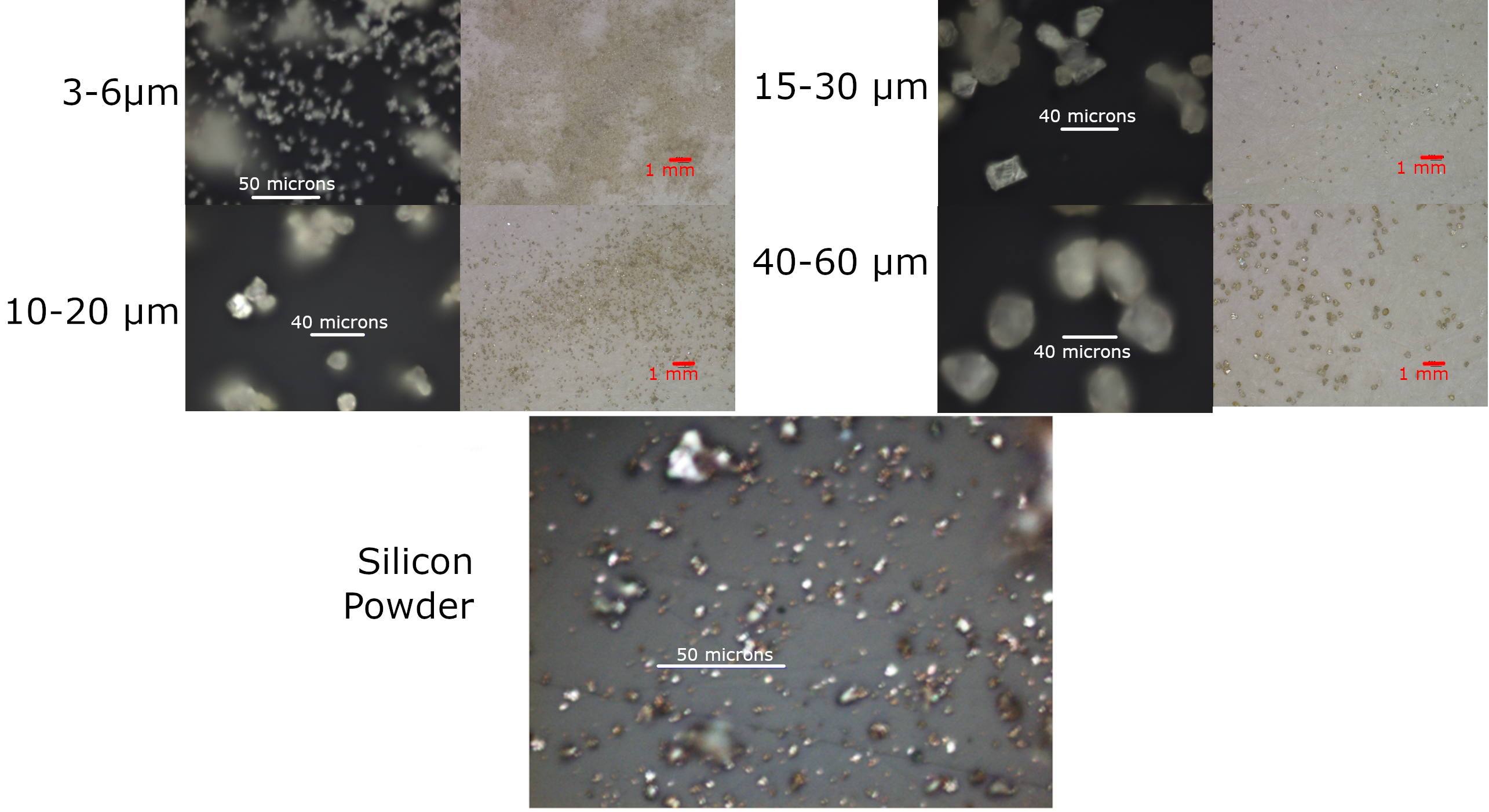}
    \caption{Optical microscope images of some of the diamond and silicon powder samples. \textit{Top:} Diamond size ranges represent the approximate three standard deviation size points, per the manufacturer. \textit{Bottom:} Even after sieving, in-house made silicon powder contained a wide size distribution with many smaller particles.}
    \label{fig:diamond_particles}
\end{figure}

Figure \ref{fig:mie_model} shows a comparison of transmission spectra between earlier modeling and fabrication efforts and more recent ones. Ultimately commercial, off-the-shelf industrial diamond particles were chosen as scattering media. We selected powders of 1-3, 3-6, 6-12, 10-20, 15-30, 20-40, and 40-60 micron distributions. All powders are from the PUREON Microdiamant MONO-ECO monocrystalline synthetic diamond. Per the manufacturer, particle size distributions are approximately Gaussian, with the end ranges representing the approximate three standard deviation limits for particle diameter. Table \ref{powders} shows a sub-sample of particle size distribution data. The particle size distributions are much more tightly controlled than the silicon powder made in-house. Diamond powder composition, particle sizes and shapes were later analyzed with optical microscopy, particle size analysis, and Raman spectroscopy. Figure \ref{fig:diamond_particles} shows images of the diamond particles and the silicon particles. 

\begin{table}[h]
\centering
\begin{tabular}{ |c|c| } 
 \hline
 \textbf{Powder Size Range} &  \textbf{Median Size} \\ 
 \hline
 3-6\,$\mu$m & 4.2\,$\mu$m \\
 \hline\
 6-12\,$\mu$m & 9\,$\mu$m \\
 \hline
 15-30\,$\mu$m & 22.5\,$\mu$m \\
 \hline
 40-60\,$\mu$m & 47\,$\mu$m \\
 \hline
\end{tabular}
\caption{Diamond particle size distributions and median sizes. All diamond powders are from the PUREON Microdiamant MONO-ECO monocrystalline synthetic diamond. Median particle sizes all fall well within the stated specifications from the manufacturer. }
\label{powders}
\end{table}

\begin{figure}[h]
    \centering
    \includegraphics[width = 0.49\textwidth]{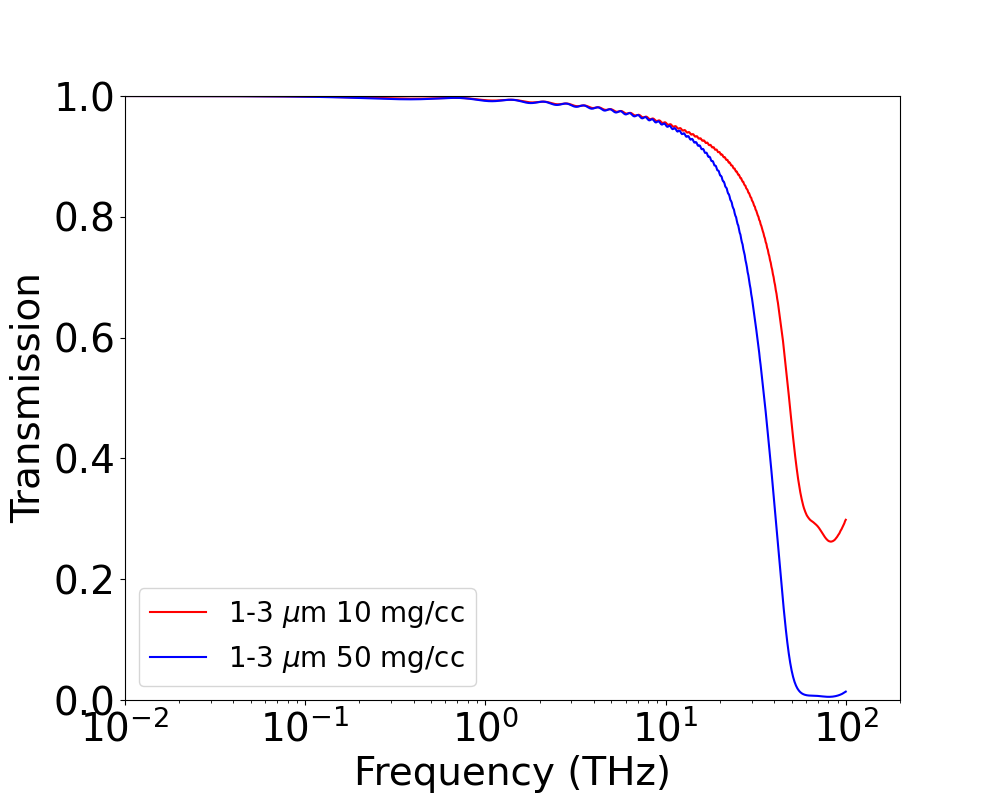}
    \includegraphics[width = 0.49\textwidth]{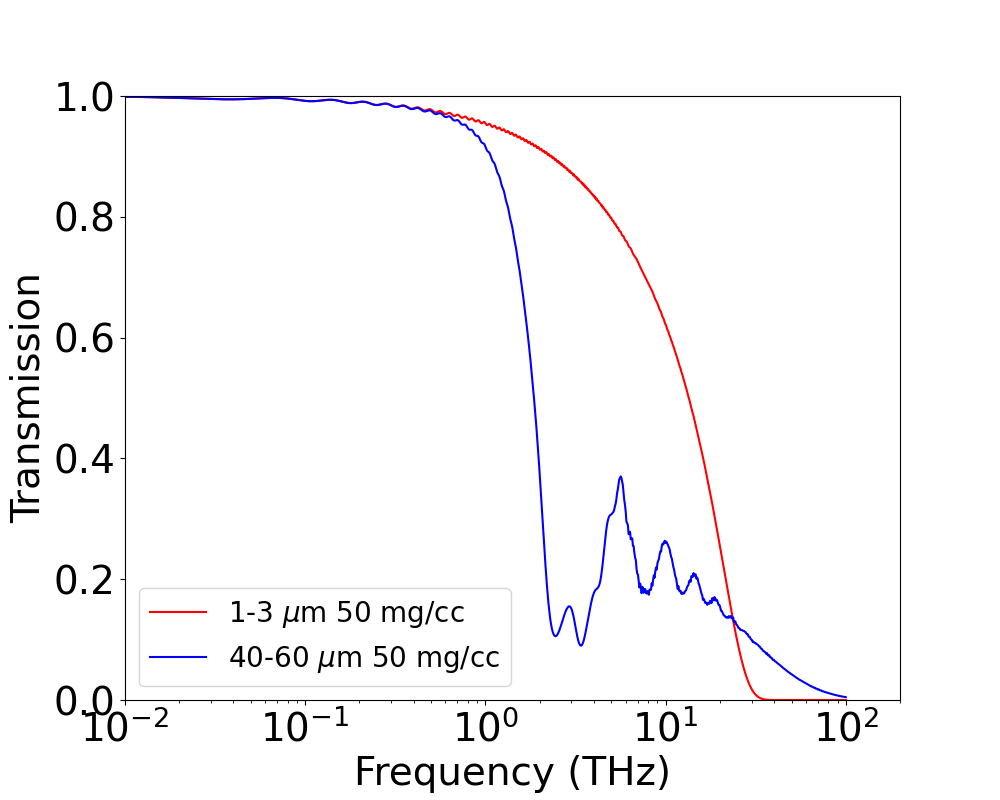}
    \caption{Demonstration of the effect of scattering particle loading density and radius on filter cut-off frequency. \textit{Left:} Comparison of Mie scattering model for different loading densities of the same particle size distribution. Both curves show the transmission of a 0.2\,mm thick filter filter with 1-3\,$\mu$m diamond particles but the red curve is for 10\,mg/cc density and the blue curve is for 50\,mg/cc. \textit{Right:} Comparison of Mie scattering model for equal loading densities of different particle size distributions. Both curves show the transmission of a 2\,mm thick filter filter with 50\,mg/cc of diamond particles but the red curve is for 1-3\,$\mu$m and the blue curve is for 40-60\,$\mu$m.  }
    \label{fig:comparison}
\end{figure}

The diamond particle size and loading density is controlled to tune the filter to the desired cutoff frequency. Particle size has the strongest effect on cutoff frequency. Larger particles have a lower cut-off frequency and smaller particles have a higher cut-off frequency. Polyimide aerogel filters have been fabricated with a range of scatterer loading densities ranging from 20\, mg/cc up to 100\, mg/cc. For a given particle size distribution, increasing the loading density creates a filter with a lower cut-off frequency and vice versa. For a given loading density, smaller particles produce a stronger cutoff than larger particles because scattering cross-section goes as the square of the particle radius (area $\propto r^2$) and mass density goes as the cube of particle radius (volume $\propto r^3$).  Figure \ref{fig:comparison} shows the effects of varying particle radius and density on the predicted filter cutoff frequency. Multiple particle size distributions and densities can be mixed together, e.g. primarily using a higher density of larger particles to set the primary cut-off frequency and mixing in lower densities of smaller particles to control high frequency leakage. 

Since first reported, continued development of the Mie scattering model has yielded additional features that allow us to capture the behavior of multiple size distributions of particles as well as include the absorption and emission features of the base polymer chemistry in the transmission curve predictions. A more detailed discussion is presented in Section \ref{sec:chemistry}. Additionally, the tightly constrained diamond particle size distributions bring the predicted filter behavior in more agreement with the measured data than in Ref \citenum{Essinger-Hileman:20}. See Ref \citenum{Barlis}, for details regarding the optical measurements of film samples. Mie scattering modeling code is now also able to accommodate user defined particle size distributions, beyond basic uniform and Gaussian options. 

\subsection{Designing Filters for Specific Missions}
With a greater control of filter cut-off frequencies, in-band transmission, and knowledge of the base aerogel transmission spectrum, it is possible to design filters to meet specific mission requirements. During the development and prototyping of filters, filters for three different missions were designed, fabricated, and characterized. 

\begin{figure}[h]
    \centering
    \includegraphics[width = 0.5\textwidth]{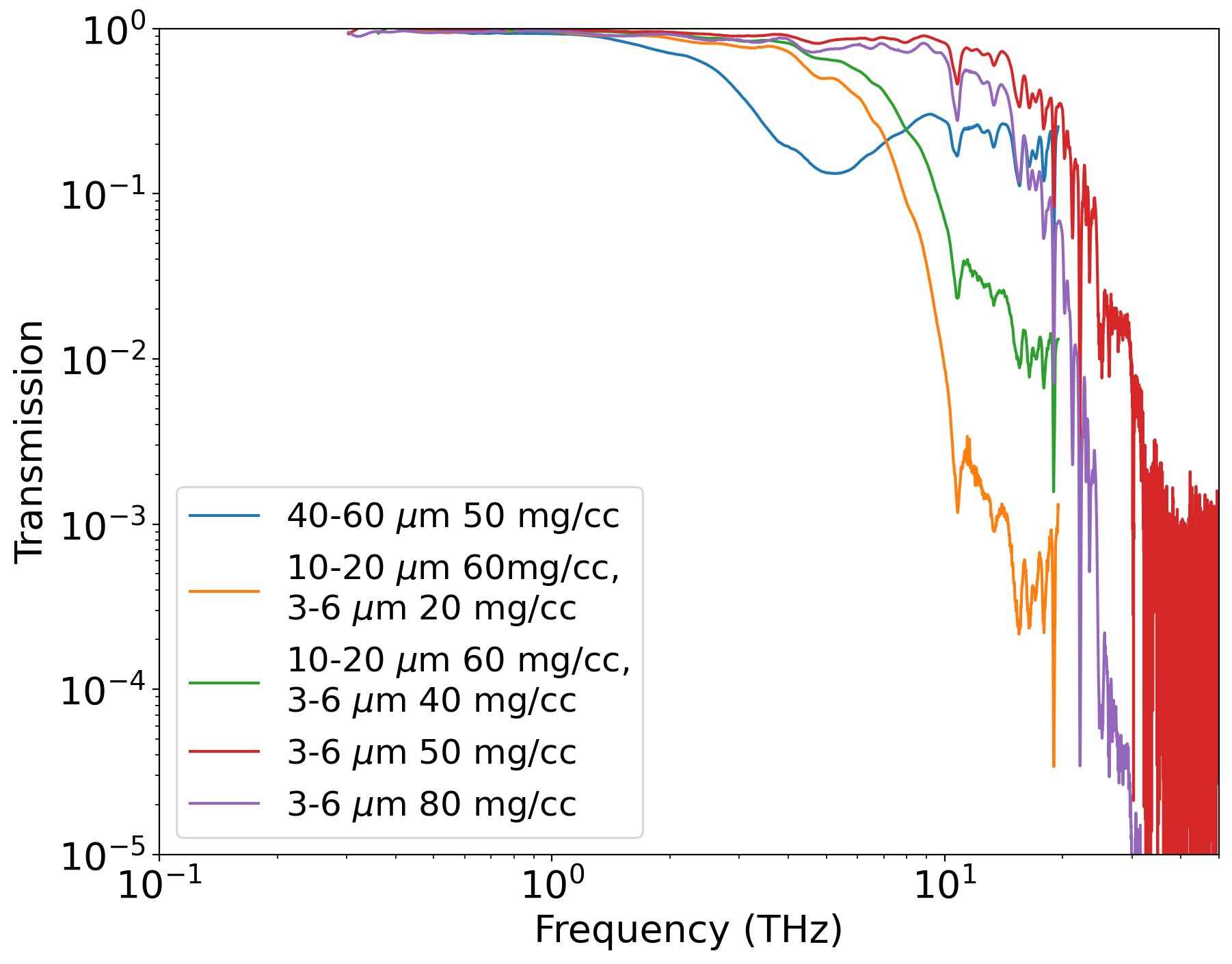}
    \includegraphics[width = 0.48\textwidth]{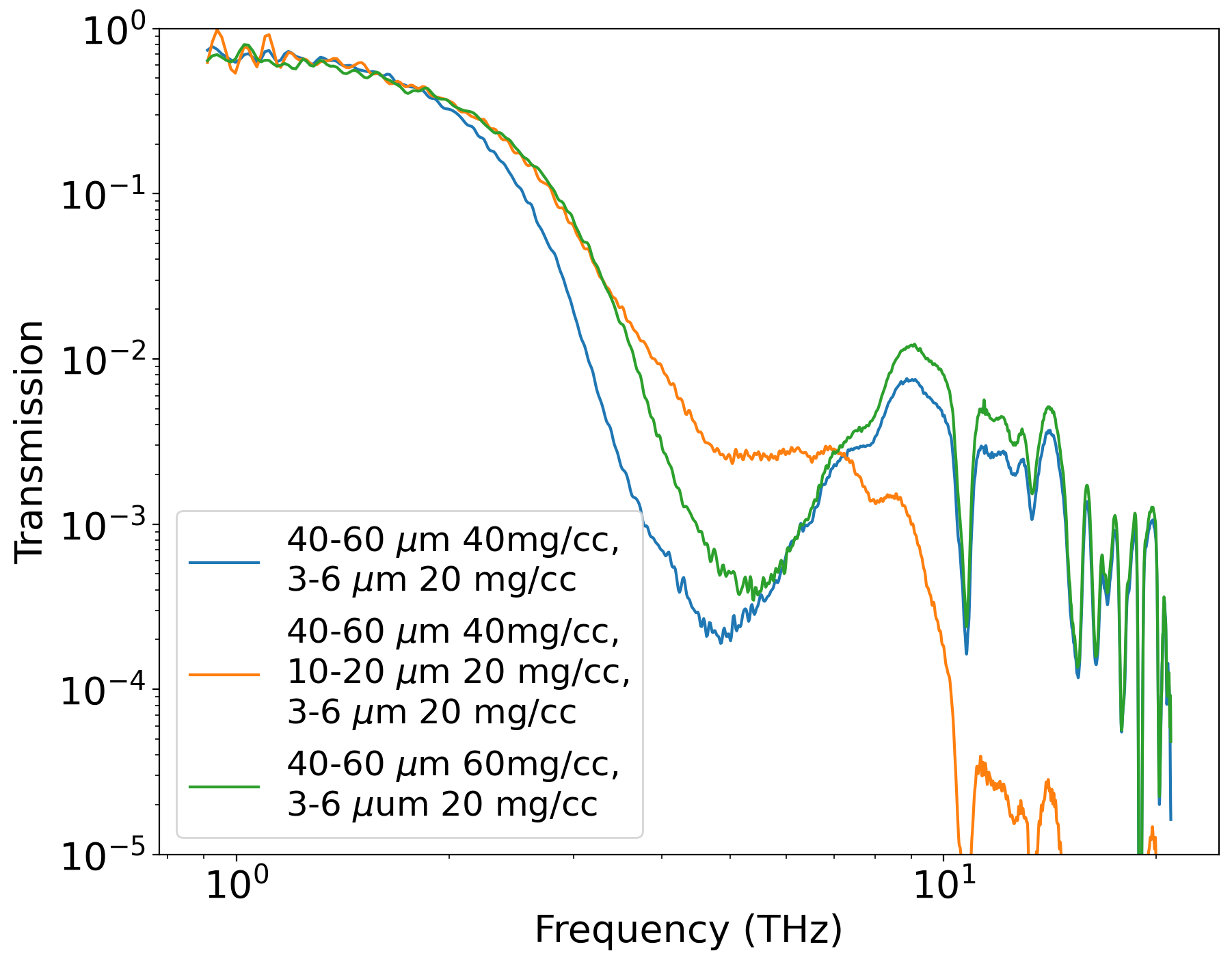}
    \caption{\textit{Left}: Transmission data for four different diamond-loaded polyimide aerogel filters for the SSOLVE mission. Because not all of the prototype filters are the same thickness, all filter transmission data were scaled to represent a 200\,$\mu$m thick sample. \textit{Right:} Transmission data for three different diamond-loaded polyimide aerogel filters for the CLASS and EXCLAIM missions. Because not all of the prototype filters are the same thickness, all filter transmission data were scaled to represent a 1\,mm thick sample. CLASS is a ground-based CMB polarimeter observes the polarized microwave sky from 33-2340\,GHz, and EXCLAIM is a balloon-borne spectrometer designed to survey at frequencies from 420-540\,GHz.}
    \label{fig:prototypes}
\end{figure}

The first experiment is the Sub-milliter Solar Observation Lunar Volatiles Experiment (SSOLVE), which will observe lines from HDO, H${}_2$O, and OH molecules in the lunar atmosphere at 23\,GHz, 500\,GHz, and 2.5\,THz, respectively\cite{SSOLVE}. SSOLVE requires good rejection of higher frequency THz light but greater than 90\% transmission at 2.5 THz. Figure \ref{fig:prototypes} shows a comparison of different diamond-loaded polyimide aerogel designed to meet the SSOLVE requirements. 

Filters have also been designed and fabricated to meet the needs of experiments that observe in the microwave regime, like the Cosmology Large Angular Scale Surveyor (CLASS), a cosmic microwave background (CMB) polarimeter designed to observe from the ground from 33-234\,GHz\cite{CLASS}. Filters designed for use in microwave observations are also being prototyped for the upcoming Experiment for Cryogenic Large-Aperture Intensity Mapping (EXCLAIM), which is a balloon-borne spectrometer that will carry out a line intensity mapping survey at frequencies from 420-540\,GHz, targeting carbon monoxide and ionized cabron emissions from galaxies in redshift slices covering $z$ = 0 - 3.5 \cite{EXCLAIM}. Figure \ref{fig:prototypes} shows several different prototype filters for these two missions.

\section{Exploring Aerogel Polymer Chemistries}
\label{sec:chemistry}
Scattering aerogel filters have been manufactured with silica and polyimide base aerogels, as well as with silicon and diamond scattering particles\cite{Essinger-Hileman:20}. As demonstrated in Ref.~\citenum{guo2}, polyimide aerogels synthesized with different diamines produce backbone structures with varying mechanical properties. However, different backbone structures may also vary the in-band transmission for infrared blocking applications. Initially, two formulations were identified for further investigation:
\begin{itemize}
    \item 4,4'-Bis (4-aminophenoxy) biphenyl (BAPB) (25mol\%); with 2,2'-Dimethylbenzidine (DMBZ)(75mol\%); 3,3',4,4'-biphenyltetracarboxylic dianhydride (BPDA); and 1,3,5- triaminophenoxybenzene (TAB), aka the ``BAPB-based polyimide'' 
    \item 1,12-dodecyldiamine (DADD)(40 mol\%); with DMBZ (60 mol\%); BPDA; and TAB, aka the ``DADD-based polyimide''.
\end{itemize}

\begin{figure}[h]
    \centering
    \includegraphics[width = 0.49\textwidth]{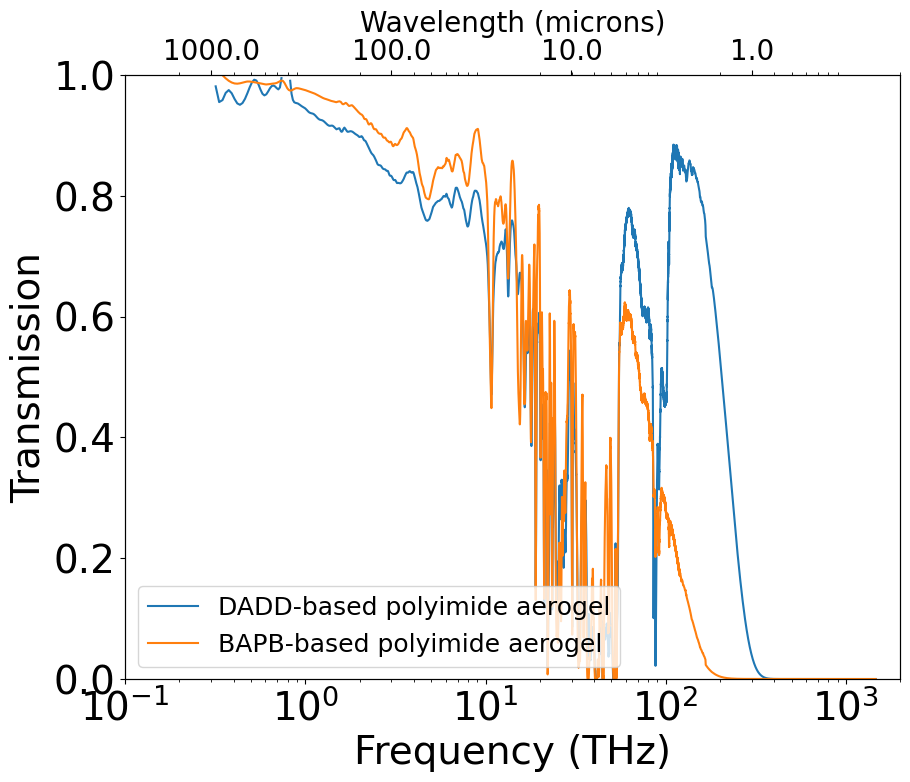}
    \includegraphics[width = 0.49\textwidth]{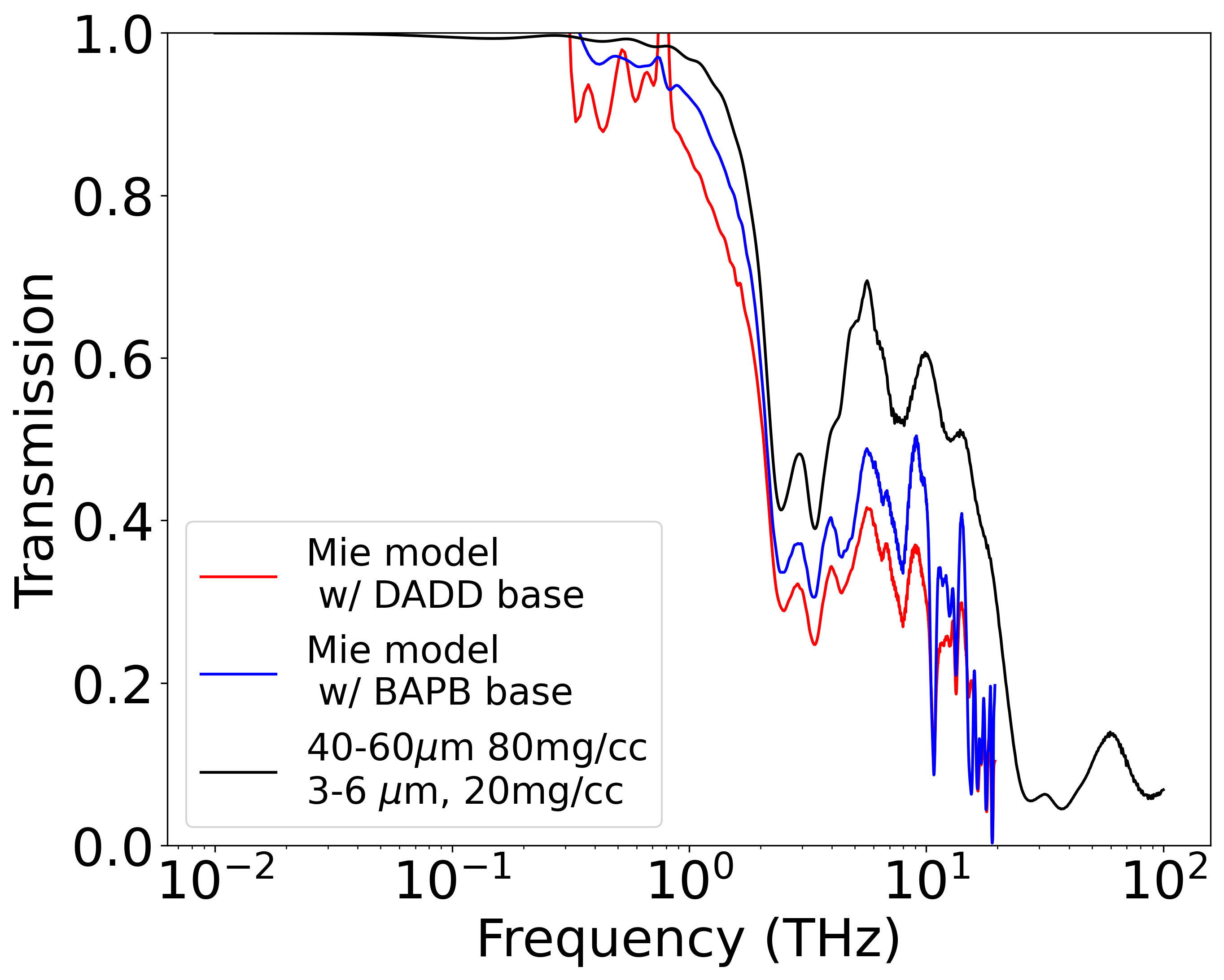}
    \caption{\textit{Left:} Comparison of the transmission spectra of two different unloaded polymimde aerogels. BAPB-based formulation is BAPB(25mol\%)+DMBZ (75mol\%)/BPDA/TAB, while the DADD-based formulation is DADD(40 mol\%)+DMBZ (60mol\%)/BPDA/TAB. Transmission spectra are scaled to represent 200 micron thick films. \textit{Right:} Model filter transmission spectra incorporating two different polyimide backbone formulations in order to allow better prediction of cutoff frequency and in-band transmission. The black curve is the Mie model alone, while the red and blue curves are the Mie model incorporating the base aerogel properties. Transmission spectra are for a 0.5\,mm thick film.}
    \label{fig:backbone_comparison}
\end{figure}

As shown in Figure \ref{fig:backbone_comparison}, the two different polymers show similar absorption features around 10-20 THz, however the BAPB formulation shows measurably higher low-frequency transmission, so this formulation was used for the vast majority of the prototype filters. 

Measuring the optical performance of the different polyimide backbone formulations also enabled the incorporation of the base aerogel properties into the Mie scattering model. Including the dielectric properties of the base aerogel into the model creates more realistic predictions of filter cutoff frequency as well as in-band transmission. Figure \ref{fig:backbone_comparison} shows a model filter with the DADD-based and BAPB-based polymers incorporated. Other polyimide polymer formulations were also explored, including 7wt\% pyrometallic dianhydride (PMDA)+DMBZ/ 1,3,5-benzenetricarbonyl trichloride (BTC) and 9wt\% PMDA+DMBZ/BTC films. Preliminary results suggest the 7wt\% PMDA+DMBZ/BTC polyimide formulation to have better in-band transmission than the BAPB-based formula. During the continued development of the polyimide-based aerogels, efforts were made to identify different polymers that might provide better in-band transmission and reduced absorption. Cyclic olefin copolymer (COC) plastics have been identified as promising candidates for aerogel substrates. The COC polymers were chosen for their high in-band transmission and reduced absorption\cite{Topas} compared to the polyimide formulations. Prototype unloaded and diamond-loaded COC aerogel filters are being manufactured. 

\section{Thermal modeling}
\label{sec:thermal}
The performance of the aerogel filters in a cryogenic receiver will ultimately determine their utility in future missions. Filter equilibrium temperatures and subsequent conductive and radiative thermal loading are key benchmarks. Thermal finite element models of the polyimide aerogel scattering filters were constructed with COMSOL Multiphysics software. The CLASS receiver filter geometry was used as a representative filter implementation. The filter stack consists of three vertically stacked filters over two filter stages, at 60\,K and 4\,K respectively. The model calculates the background infrared loading on the 4\,K stage after loading from a 250\,K surface passes through the upper filters. The polyimide aerogel filter performance was compared with model alumina and polytetrafluoroethylene (PTFE) filters. The thermal conductivities of each material are described in Table \ref{tab:filters}. 

\begin{table}[]
    \centering
    \begin{tabular}{|c|c|}
        \hline
         \textbf{Material} & \textbf{Thermal Conductivity $[\frac{\text{W}}{\text{m\,K}}]$}\\
         \hline
         Diamond-loaded polyimide aerogel & 0.024 \\
         \hline
         Alumina & 27 \\
         \hline
         PTFE & 0.24 \\
         \hline 
    \end{tabular}
    \caption{Summary of material properties used in COMSOL thermal finite element analysis models. Alumina and PTFE are commonly used materials for infrared rejection in cryogenic telescope receivers in the sub-millimeter and microwave.}
    \label{tab:filters}
\end{table}

\begin{figure}
    \centering
    \includegraphics[width = 0.4\textwidth]{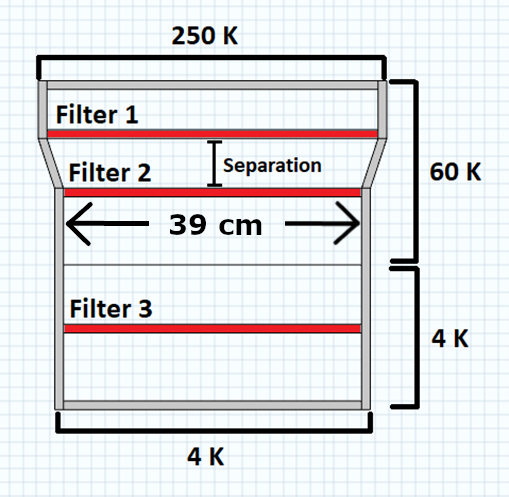}
    \caption{Schematic of the three filter stack inside of the CLASS receiver. The first two filters are both mounted at the 60\,K stage but with a variable gap between them. The third filter is mounted to the 4\,K stage. The 250\,K surface represents the backside of the stage above the 60\,K. As pictured, the two 60\,K filters are 43\,mm apart.}
    \label{fig:filter_stack}
\end{figure}

The first two filters are mounted to the 60\,K stage and the side wall of the optics tube around the top two filters is held at 60\,K. The third filter was mounted to the 4\,K stage and the sidewall along the bottom filter as well as the wall between the last filter and the focal plane were held at 4\,K. The surface above the top of the first filter is held at 250\,K to simulate the radiation incoming from the stage that sits above the 60\,K stage in the CLASS receiver. The separation of filters within the 60K stage of the receiver can be varied to minimize the background radiation that reached the 4\,K stage. The geometry of the filters from the COMSOL model is shown in Figure \ref{fig:filter_stack}.

\begin{figure}
    \centering
    \includegraphics[width = \textwidth]{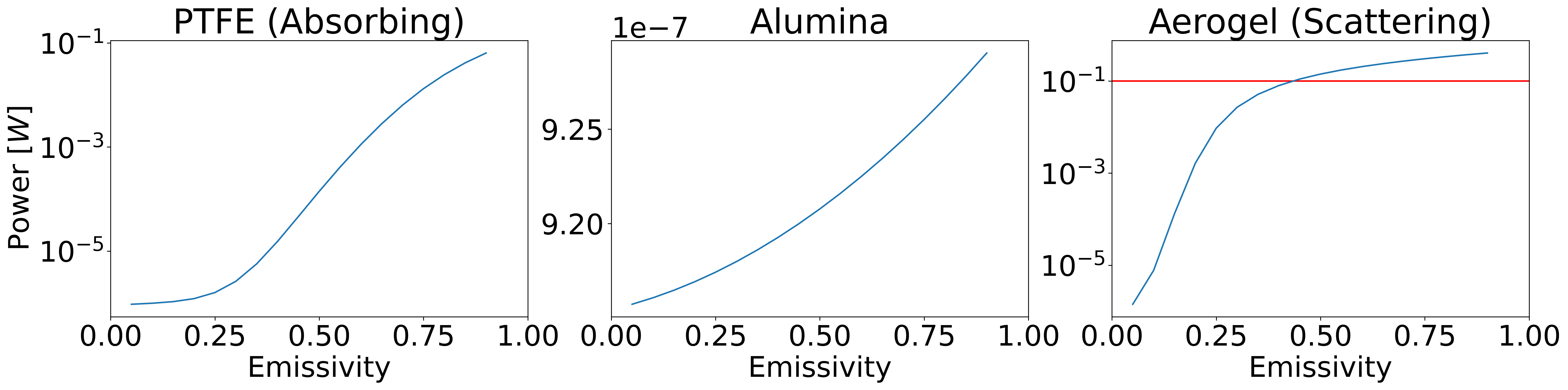}
    \caption{Total thermal load on the third filter for three different filter substrates and a range of emissivity values. The two 60\,K filters are spaced 43\,mm apart. The left plot is for absorbing PTFE filters, the middle plot for alumina filters, and the right plot is for the diamond loaded polyimide aerogel filters. Three alumina filters provide the greatest rejection of background loading power, but with lower in-band transmission and greater cost and complexity. Additionally, most implementations of alumina filters do not use a three filter stack. The simulation results depend strongly on the emissivity of the aerogel filters. CLASS requires less than 100\,mW of load, indicated by the red horizontal line in the right plot.}
    \label{fig:emissivity_sweep}
\end{figure}

The COMSOL radiation module was used to estimate the total thermal load on the 4\,K stage. In addition to the thermal loading from electromagnetic radiation, the COMSOL model also includes the thermal loading due to conduction through the parts of the filter stack in physical contact with other components of the receiver. The emissivity of the aerogel, alumina, and PTFE filters was swept parametrically from $\epsilon$ = 0.05 to 0.9, in order to predict filter performance across a wide range of physical parameters. A total equilibrium thermal load on the third filter was calculated for each emissivity. Figure \ref{fig:emissivity_sweep} shows the comparison of the performance between the three different filter substrates. The two 60\,K filters are held at the nominal CLASS receiver spacing of 43\,mm. A stack of three alumina filters shows the most power rejection by several orders of magnitude. The results are a bit misleading, however, because most implementations of alumina filters do not use a stack of three because of their lower in-band transmission. Alumina filteras also have increased cost and more complex AR-coating requirements, due to the high index of refraction, n$\simeq 3.1$. The performance of the aerogel filters depends strongly on the emissivity, but they meet the CLASS thermal loading requirement of less than 100\,mW as long as the emissivity is below 0.5. In contrast, alumina and PTFE filters are high emissivity by design, with expected emissivities of 0.75 and 0.97 respectively, set by reflectivity from the surface. 

A second simulation of aerogel filters was run to determine the optimal filter spacing of the 60\,K stack. Total thermal load on the 4\,K filter was calculated for spacing ranging from 27\,mm to 120\,mm at emissivities of 0.1, 0.5, and 0.9. The current CLASS design has a spacing of 43\,mm. Maximal filter spacing minimizes thermal loading in all cases, by approximately a factor of 5. The data suggest that spacing the 60\,K filters as distant apart as is reasonable is optimal. Cryogenic testing to verify the predicted thermal performance of the aerogel scattering filters is expected to begin in the fourth quarter of 2022.

\section{Concluding Remarks}
Diamond-loaded polyimide aerogel scattering filters are a promising emerging filter technology for use in far-infrared, sub-millimeter, and microwave astrophysics, cosmology, and planetary science missions. Prototype filters demonstrate excellent out-of-band rejection, high in-band transmission, and tunable cut-off frequencies. Filters are being manufactured in large diameters (larger than 40\,cm) and cryogenic testing is underway to verify their performance. In addition to polyimide aerogel filters, other polymeric aerogel filters are being explored and will potentially yield better in-band transmission with reduced in and out-of-band absorption.

\acknowledgments 
 
The authors would like to thank Aerogel Technologies, LLC, for their assistance in manufacturing film prototypes while NASA facilities were closed due to the COVID-19 pandemic. The material is based upon work supported by NASA under award number 80GSFC21M0002. 
\bibliography{report} 
\bibliographystyle{spiebib} 

\end{document}